\documentclass{article}

\begin{document}

\begin{center}
\bigskip Model of Extended Mechanics
and Non-Local Hidden Variables for Quantum Theory

\textit{Timur F. Kamalov}

Physics Department\\[0pt]
Moscow State Open University\\[0pt]
107996 Moscow, 22, P. Korchagin str., Russia\\[0pt]
E-mail: ykamalov@rambler.ru\\[0pt]
www.TimKamalov.narod.ru\\[0pt]
\end{center}

\begin{abstract}
Newtonian physics is describes macro-objects sufficiently well,
however it does not describe microobjects. A model of Extended
Mechanics for Quantum Theory is based on an axiomatic
generalization of Newtonian classical laws to arbitrary reference
frames postulating the description of body dynamics by
differential equations with higher derivatives of coordinates with
respect to time but not only of second order ones and follows from
Mach principle. In that case the Lagrangian
$L(t,q,\dot{q},\ddot{q},...,\dot {q}^{(n)},...)$ depends on higher
derivatives of coordinates with respect to time. The kinematic
state of a body is considered to be defined if n-th derivative of
the body coordinate with respect to time is a constant (i.e.
finite). First, kinematic state of a free body is postulated to
invariable in an arbitrary reference frame. Second, if the
kinematic invariant of the reference frame is the n-th order
derivative of coordinate with respect to time, then the body
dynamics is describes by a 2n-th order differential equation. For
example, in a uniformly accelerated reference frame all free
particles have the same acceleration equal to the reference frame
invariant, i.e. reference frame acceleration. These bodies are
described by third-order differential equation in a uniformly
accelerated reference frame.
\end{abstract}



\vspace{5mm}

\section{Introduction}
 Classical Newtonian mechanics is essentially the simplest way of
mechanical system description with second-order differential
equations, when higher order time derivatives of coordinates can
be neglected. The extended model of mechanics with higher time
derivatives of coordinates is based on generalization of Newton's
classical axiomatics onto arbitrary reference frames (both
inertial and non-inertial ones) with body dynamics being described
with higher order differential equations. Newton's Laws,
constituting, from the mathematical viewpoint, the axiomatics of
classical physics, actually postulate the assertion that the
equations describing the dynamics of bodies in inertial frames are
second-order differential equations. However, the actual
time-space is almost without exception non-inertial, as it is
almost without exception that there exist (at least weak) fields,
waves, or forces perturbing an ideal inertial frame. It
corresponds to Mach's principle [1] with general statement "Local
physical laws are determined by the large-scale structure of the
universe." Non-inertial nature of the actual time-space is also
supported by observations of the practical astronomy that
expansion of the reality occurs with an acceleration. In other
words, actually any real reference frame is a non-inertial one;
and such physical reality can be described with a differential
equation with time derivatives of coordinates of the order
exceeding two, which play the role of additional variables. This
is evidently beyond the scope of Newtonian axiomatics. Aristotle's
physics considered velocity to be proportional to the applied
force, hence the body dynamics was described by first derivative
differential equation. Newtonian axiomaics postulates reference
frames, where a free body maintains the constant velocity of
translational motion. In this case the body dynamics is described
with a second order differential equation, with acceleration being
proportional to force [2]. This corresponds to the Lagrangian
depending on coordinates and their first derivatives (velocities)
of the body, and Euler-Lagrange equation resulting from the
principle of the least action. This model of the physical reality
describes macrocosm fairly good, but it fails to describe micro
particles. Both Newtonian axiomatics and the Second Law of Newton
are invalid in microcosm. Only averaged values of observable
physical quantities yield in the microcosm the approximate analog
of the Second Law of Newton; this is the so-called Ehrenfest's
theorem. The Ehrenfest's equation yields the averaged, rather than
precise, ration between the second time derivative of coordinate
and the force, while to describe the scatter of quantum
observables the probability theory apparatus is required. As the
Newtonian dynamics is restricted to the second order derivatives,
while micro-objects must be described with equations with
additional variables, tending Planck's constant to zero
corresponds to neglecting these variables. Hence, offering the
model of extended Newtonian dynamics, we consider classical and
quantum theories with additional variables, describing the body
dynamics with higher order differential equations. In our model
the Lagrangian shall be considered depending not only on
coordinates and their first time derivatives, but also on
higher-order time derivatives of coordinates. Classical dynamics
of test particle motion with higher-order time derivatives of
coordinates was first described in 1850 by M.Ostrogradskii [3] and
is known as Ostrogradskii's Canonical Formalism. Being a
mathematician, M. Ostrogradskii considered coordinate systems
rather than reference frames. This is just the case corresponding
to a real reference frame comprising both inertial and
non-inertial reference frames. In a general case, the Lagrangian
takes on the form $(n\rightarrow \infty)$

\begin{equation}
L=L(t,q,\dot{q},\ddot{q},...,\dot{q}^{(n)}).
\end{equation}

\section{Theory of Extended Mechanics}
Let us consider in more detail this precise description of the
dynamics of body motion, taking into account of real reference
frames. To describe the extended dynamics of a body in an any
coordinate system (corresponding to arbitrary reference frame) let
us introduce concepts of kinematic state and kinematic invariant
of an arbitrary reference frame.

\textbf{Definition}: Kinematic state of a body is set by $n$-th
time derivative of coordinate. The kinematic state of the body is
defined
provided the $n$-th time derivative of body coordinate is zero, the $(n-1)$%
-th time derivative of body coordinate being constant. In other
words, we consider the kinematic state of the body defined if
$(n-1)$-th time derivative of body coordinate is finite. Let us
note that a reference frame performing harmonic oscillations with
respect to an inertial reference frame does not possess any
definite kinematic state. Considering the dynamics of particles in
arbitrary reference frames, we suggest the following two
postulates.

\textbf{Postulate 1.} Kinematic state of a free body is
invariable. This means that if the $n$-th time derivative of a
free body coordinate is zero, the $(n-1)$-th time derivative of
body coordinate is constant. That is,

\begin{equation}
\frac{d^{n}q}{dt^{n}}=0,\frac{d^{n-1}q}{dt^{n-1}}=const.
\end{equation}

In the extended model of dynamics, conversion from a reference
frame to another one will be defined as:
\begin{eqnarray}
q^{\prime } &=&q_{0}+\dot{q}t+\frac{1}{2!}\ddot{q}t^{2}+...+\frac{1}{n!}%
\dot {q}^{(n)}t^{n}.
\end{eqnarray}

\textbf{Postulate 2.} If the kinematic invariant of a reference frame is $n$%
-th time derivative of body coordinate, then the body dynamics is
described with the differential equation of the order $2n$:

\begin{equation}
\alpha _{2n}\dot {q}^{(2n)}+...+\alpha _{0}q=F(t,q,\dot{q},\ddot{q},...,%
\dot {q}^{(n)}).
\end{equation}

This means that the Lagrangian depends on $n$-th time derivative
of coordinate, so variation when applying the least action
principle will yield the order higher by a unity. Therefore, the
dynamics of a free body in a reference frame with $n$-th order
derivative being invariant shall be described with a differential
equation of the order $2n$. To consider dynamics of a body with an
observer in an arbitrary coordinate system, varying the action
function for $n$-th order kinematic invariant, we obtain the
equation of the order $2n$:

\begin{equation}
\delta S=\delta \int L(t,\dot{q^{\prime }},q^{\prime })dt =\int
\sum_{n=0}^{N}(-1)^{n}\frac{d^{n}}{dt^{n}}\frac{\partial L}{\partial \dot{q}%
^{(n)}}\delta \dot{q}^{(n)}dt=0.
\end{equation}

Expanding into Taylor's series the function $q=q(t)$ yields:

\begin{equation}
q=q_{0}+\dot{q}t+\frac{1}{2!}\ddot{q}t^{2}+...+\frac{1}{n!}\dot{q}%
^{(n)}t^{n}.
\end{equation}

It is well known that the kinematic equation in inertial reference
frames of Newtonian physics contains the second time derivative of
coordinate, that is, acceleration:

\begin{equation}
q_{Newton}=q_{0}+vt+\frac{1}{2}at^{2}.
\end{equation}
Let us denote the additional terms with higher derivatives as

\begin{equation}
q_{r}=\frac{1}{3!}\dot{q}^{(3)}t^{3}+...+\frac{1}{n!}\dot{q}^{(n)}t^{n}.
\end{equation}

Then
\begin{equation}
q=q_{newton}+q_{r}.
\end{equation}

In our case, the discrepancy between descriptions of the two
models is the difference between the description of test particles
in the model of Extended Mechanics with Lagrangian $L(t,q,\dot{q},\ddot{q},...,%
\dot {q}^{(n)},...)$ and Newtonian dynamics in inertial reference
frames with the Lagrangian $L(t,q,\dot{q})$:

\begin{equation}
\int [L(t,q,\dot{q},\ddot{q},...,\dot
{q}^{(n)})-L(t,q,\dot{q})]dt=h,
\end{equation}
$h$ being the discrepancy (error) between descriptions by the two
models. Comparing this value with the uncertainty of measurement
in inertial reference frames, expressed by the Heisenberg
uncertainty relation, the equation (10) can be rewritten as
\begin{equation}
S(t,q,\dot{q},...\dot{q}^{(n)})-S(t,q,\dot{q})=h.
\end{equation}

In the classical mechanics, in inertial reference frames, the
Lagrangian depends only on the coordinates and their first time
derivatives. In the Extended Mechanics, in real reference frames,
the Lagrangian depends not only on the coordinates and their first
time derivatives, but also on their higher derivatives. Applying
the least action principle [4], we obtain Euler-Lagrange equation
for the Extended Mechanics:

\begin{equation}
\sum_{n=0}^{N}(-1)^{n}\frac{d^{n}}{dt^{n}}\frac{\partial L}{\partial \dot{q}%
^{(n)}}=0,
\end{equation}

or
\begin{equation}
\frac{\partial L}{\partial q}-\frac{d}{dt}\frac{\partial L}{\partial \dot{q}}%
+\frac{d^{2}}{dt^{2}}\frac{\partial L}{\partial \ddot{q}}-...+(-1)^{N}\frac{%
d^{N}}{dt^{N}}\frac{\partial L}{\partial \dot{q}^{(N)}}=0.
\end{equation}
The Lagrangian will be expressed through quadratic functions of
variables:
\begin{equation}
L=kq^{2}-k_{1}\dot{q}^{2}+k_{2}\ddot{q}^{2}-...+(-1)^{\alpha }k_{\alpha }%
\dot{q}^{(\alpha )2}=\sum_{\alpha =0}^{\infty }(-1)^{\alpha }k_{\alpha }\dot{%
q}^{(\alpha )2}.
\end{equation}
For our case, the action function will be:
\begin{equation}
S=q\frac{\partial L}{\partial q}-\dot{q}\frac{\partial L}{\partial \dot{q}}%
+...+(-1)^{\alpha }\dot{q}^{(\alpha )}\frac{\partial \dot{L}^{(\alpha )}}{%
\partial \dot{q}^{(\alpha )}}+...=\sum_{\alpha =0}^{\infty }(-1)^{\alpha }%
\dot{q}^{(\alpha )}\frac{d^{\alpha }}{dt^{\alpha }}\frac{\partial L}{%
\partial \dot{q}^{(\alpha )}}.
\end{equation}

Or
\begin{equation}
S=2kq^{2}-2k_{1}\dot{q}^{2}+2k_{2}\ddot{q}^{2}+...+2k_{\alpha }\dot{q}%
^{(\alpha )2}=2\sum_{\alpha =0}^{\infty }(-1)^{\alpha }k_{\alpha }\dot{q}%
^{(\alpha )2}.
\end{equation}

Introducing the notation

\begin{center}
\begin{equation}
F=\frac{\partial L}{\partial q},p=\frac{\partial L}{\partial
\dot{q}}
\end{equation}%
\begin{equation}
F^{2}=\frac{\partial L}{\partial \ddot{q}},p^{3}=\frac{\partial
L}{\partial \dot{q}^{(3)}}
\end{equation}%
\begin{equation}
F^{4}=\frac{\partial L}{\partial \dot{q}^{(4)}},p^{5}=\frac{\partial L}{%
\partial \dot{q}^{(5)}}
\end{equation}%
.....\\[0pt]
\begin{equation}
F^{2n}=\frac{\partial L}{\partial \dot{q}^{(2n)}},p^{2n+1}=\frac{\partial L}{%
\partial \dot{q}^{(2n+1)}},
\end{equation}
\end{center}

we obtain the description of inertial forces for Extended
Mechanics. The value of the resulting force accounting for
inertial forces can be expressed through momentums and their
derivatives, expressing the Second Law of Newton for the extended
Newtonian dynamics model:
\begin{equation}
F-\frac{dp}{dt}+\frac{d^{2}}{dt^{2}}(F^{2}-\frac{dp^{3}}{dt})+\frac{d^{4}}{%
dt^{4}}(F^{4}-\frac{dp^{5}}{dt})+...\frac{d^{n}}{dt^{n}}(F^{n}-\frac{dp^{n+1}%
}{dt})=0.
\end{equation}

In other words, (21) can be written as
\begin{equation}
\sum_{n=0}^{\infty }\frac{d^{2n}}{dt^{2n}}(F^{2n}-\frac{d^{2n}p^{2n+1}}{%
dt^{2n}})=0.
\end{equation}
The action function takes on the form
\begin{equation}
S=\sum_{n=0}^{\infty }(-1)^{n}\dot{q}^{(n)}p^{n+1}=\sum_{n=0}^{N}(-1)^{n}%
\dot{q}^{(n)}\frac{\partial L}{\partial \dot{q}^{(n+1)}}.
\end{equation}

For this case, energy can be expressed as
\begin{equation}
E=\alpha _{0}q^{2}+\alpha _{1}\dot{q}^{2}+\alpha
_{2}\ddot{q}^{2}+...+\alpha _{n}\dot{q}^{(n)2}+...
\end{equation}%
Denoting the Appel's energy of acceleration [5] as $Q$, $\alpha
_{n}$ being constant factors, we obtain for kinetic energy and
potential energy, respectively,
\begin{eqnarray}
E &=&V+W+Q  \label{26} \\
V &=&\alpha _{0}q^{2}\\
W &=&\alpha _{1}\dot{q}^{2}\\
Q &=&\alpha _{2}\ddot{q}^{2}+...+\alpha _{n}\dot{q}^{(n)2}+...
\end{eqnarray}%
The Hamilton-Jacobi equation for the action function will take on
the form
\begin{equation}
-\frac{\partial S}{\partial t}=\frac{(\nabla S)^{2}}{2m}+V+Q,
\end{equation}%
The first addend in (28) is the so-called Appel's energy of
acceleration [5]. Let us compare $Q$ with the Bohm's quantum
potential [6] and complement the equation (29) with the continuity
equation. If $Q\approx \alpha _{2}\frac{\nabla ^{2}S}{m^{2}}$
(here, the value of the constant is chosen $\alpha
_{2}=\frac{i\hbar m}{2}$). Hence, in the first approximation we
obtain for the function $\psi =e^{\frac{i}{\hbar }S}$, the
Schroedinger equation\bigskip
\begin{equation}
i\hbar \frac{\partial \psi }{\partial t}=\frac{\hbar
^{2}}{2m}\nabla ^{2}\psi +V\psi .
\end{equation}

\section{Conclusions}

Our case corresponds to Lagrangian $L(t,q,\dot{q},\ddot{q},...,\dot {q}%
^{(n)},...)$, depending on coordinates, velocities and higher time
derivatives, which we call additional variables, extra addends, or
hidden variables. In arbitrary reference frames (including
non-inertial ones) additional variables (addends) appear in the
form of higher time derivatives of coordinates, which complement
both classical and quantum physics. It should be noted that these
hidden (addition) variables can be used to complement the quantum
description without violating von Neumann theorem, as this theorem
is not applied for non-linear reference frames, while the extended
mechanics assumes employing any reference frames, including
non-linear ones. For example, if we consider a spaceship with two
observers in different cabins, one can see that this system is
non-ideal, the inertial forces (or pseudo-forces) could constitute
additional variables here. In this case, superposition of the two
distributions obtained by the observers could yield a non-zero
correlation factor, though each of the two observations has a
seemingly random nature. If the fact that the reference frame is
non-inertial and hence there exist additional variables in the
form of inertial effects is ignored, then non-local correlation of
seemingly independent observations would seem surprising. This
example could visualize not only the interference of corpuscle
particles, but also the non-local character of quantum
correlations when considering the effects of entanglement.
Newtonian mechanics (i.e. without additional derivatives) work so
well in applications is valid in the framework of its
applicability with a certain accuracy. The Extended Mechanics has
a wider field of applicability, with Newtonian mechanics being its
particular case. When Newtonian mechanics is invalid the Extended
Mechanics acquires additions variables in the form of higher
derivatives. Averaging we obtain Erenfest's theorem. Introducing
higher derivatives (i.e. hidden variables) means transition to
Quantum Mechanics. Thus quantum mechanical experiments confirm the
Extended Mechanics as well. Analogies Extended Mechanics to Bohm's
mechanics confirm this viewpoint. The next question appears: is
this equation linear or not? We consider that the contribution of
non-linear composed is small and in the first approximation gives
the Schroedinger equation. The model admits non-linear
generalization. The present model of extended Newtonian dynamics
is generalize but not alternative to Newtonian Dynamics because
its extended mechanics to arbitrary reference frames. It is
physics of arbitrary reference frames. Extended Mechanics
describes the dynamics of mechanical systems for arbitrary
reference frames and not only for inertial reference frames as
Newtonian Dynamics. Newtonian Dynamics can describe non-inertial
reference frames as well introducing fiction forces. In Extended
Mechanics (Dynamics) we have fiction forces naturally and
automatically from new axiomatic and we needn't have inertial
reference frame. Model of Extended Dynamics is differs from
Newtonian Dynamics in the case of micro-objects description.

\end{document}